\documentstyle[11pt,newpasp,twoside,epsfig]{article}
\def\edcomment#1{\iffalse\marginpar{\raggedright\sl#1\/}\else\relax\fi}
\reversemarginpar

\begin{document}
\title{Kinematic Study of the Blazar 0716+714}
 \author{U. Bach, T.P. Krichbaum, E. Ros, A. Witzel \& J.A. Zensus}
\affil{\small Max-Planck-Institut f\"ur Radioastronomie, Auf dem H\"ugel 69, 53121
Bonn, Germany}
\author{S. Britzen}
\affil{\small Landessternwarte, K\"onigstuhl 17, 69117 Heidelberg, Germany Max-Planck-Institut f\"ur Radioastronomie, Bonn, Germany}
\begin{abstract}
{\small We determined the kinematics at the jet of 0716+714 from a reanalysis of multi-frequency
VLBI data (5, 8.4, 15, 22\,GHz) obtained during the last 10 years combined with
data from the literature. For this intra-day variable blazar, only a lower limit
of its distance is known ($z \geq 0.3$). We find that 0716+714 is a relatively
fast superluminal source (with a Lorentz factor of $\gamma>15$), revising
earlier results showing much slower motions. We discuss the new findings with
emphasis on the interpretation of the observed rapid radio variability. }
\end{abstract}
\vspace{-1cm}
\section{Introduction}
The BL Lac object 0716+714 is extremely variable on time-scales from hours to
months at all observed wavelengths from radio to X-ray. The redshift of 0716+714
is not yet known. However, optical imaging of the underlying galaxy provides a
lower limit to the distance of $z \geq 0.3$ (Wagner et al. 1996). In the radio
bands 0716+714 shows intraday variability (IDV) (Witzel et al. 1986; Heeschen et al.
1987). It exhibits a very flat radio spectrum, extending up to at least
350\,GHz. The variability appears to be correlated between radio and optical
(Quirrenbach et al. 1991) and over wide ranges of the electromagnetic spectrum
(Wagner et al. 1996). The simultaneous variations between X-ray, optical and
radio strongly suggest an intrinsic origin for the variability. VLBI studies
covering more than 20\,years at 5\,GHz show a core-dominated evolving jet
extending to the north, with contradicory measurements of its proper motion
ranging from 0.05\,mas\,yr$^{-1}$ to 1.1\,mas\,yr$^{-1}$ (Eckart et al.\
1986,\ 1987; Witzel et al.\ 1988; Schalinski et al.\ 1992; Gabuzda et al.\ 1998;
Tian et al.\ 2001; Jorstad et al.\ 2001). 

Our analysis includes data at 5\.GHz at four epochs from the CJF-Survey between
1992 and 1999 (Britzen et al. 1999) and a VSOP observation from 2000, data at
8.4\,GHz at two epochs from astrometric observations (Ros et al. 2001) and three
epochs of our own observations between 1994 and 1999, data at 15\,GHz at five
epochs from the VLBA 2\,cm Survey from 1994 to 2001 (Kellermann et al. 1998,
Zensus et al. 2002) and data at 22\,GHz at seven epochs from Jorstad et al.
(2001) and four epochs of our own data from 1992 to 1997. The observational
details and the data reduction will be described in detail in Bach et al. (2003
in prep.). Examples of those are presented in Fig.~1. 
\begin{figure}
\plottwo{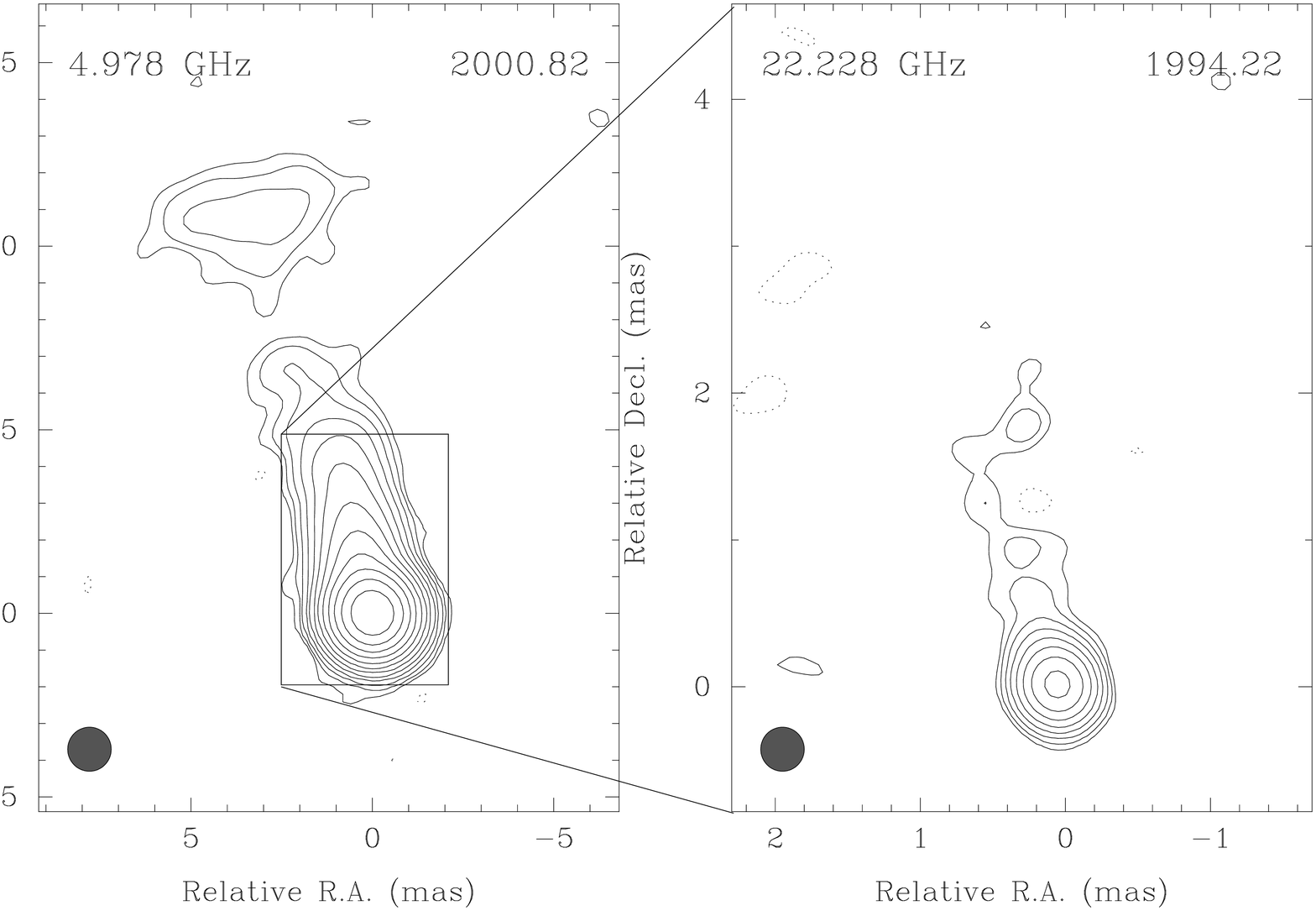}{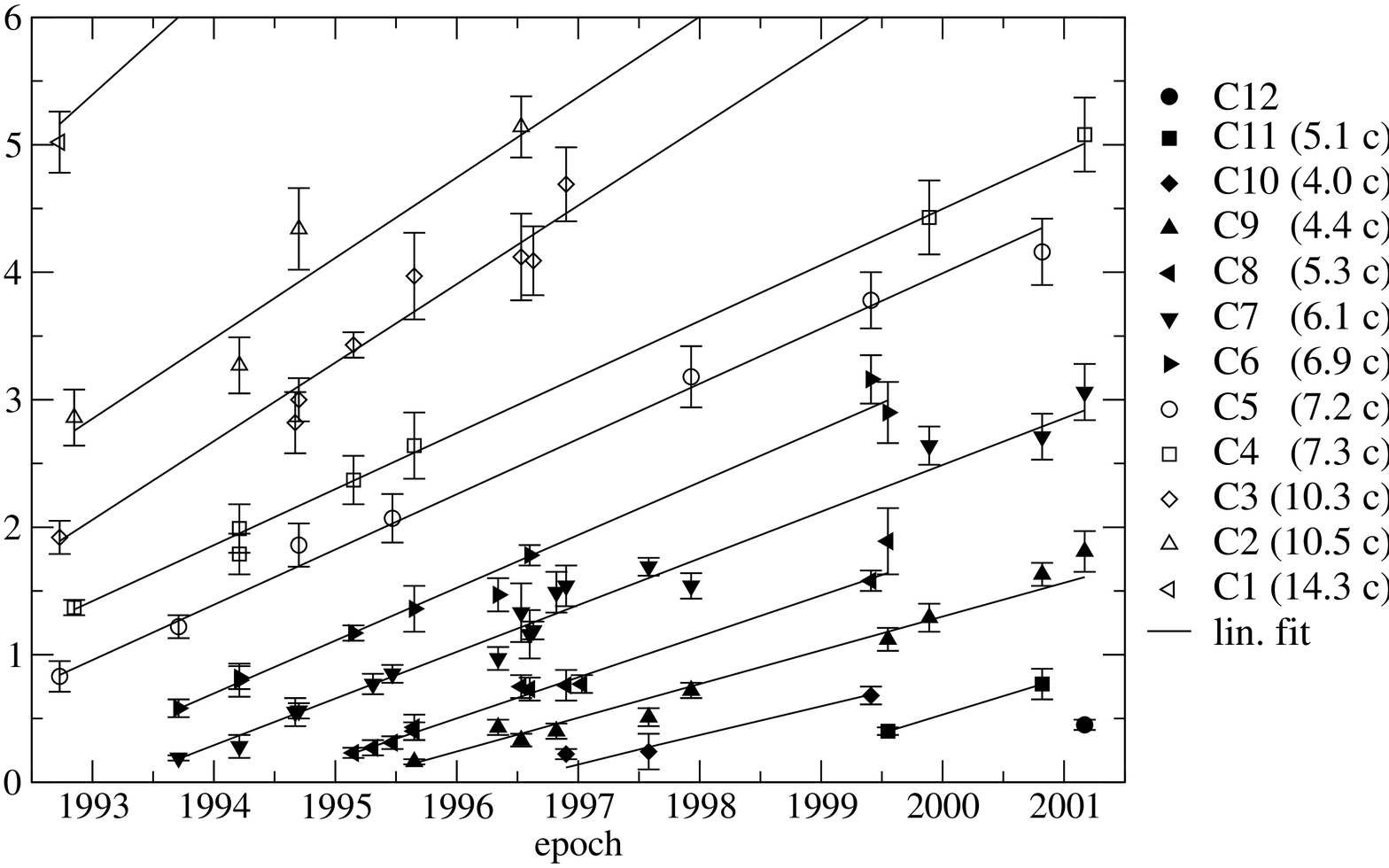}
\caption{\footnotesize Left: Contour maps at 5 and 22\,GHz at different epochs convolved with
circular beams of 1.2\,mas and 0.3\,mas respectively. Right:
The core distance of individual VLBI components derived from our model-fits at
various frequencies are plotted against time. The solid lines are linear fits to
the path of the components. The velocity derived from the fits are set in
parenthesis behind the component labels in the legend.}
\end{figure}
\section{Results and Discussion}
To investigate the kinematics in the jet of 0716+714 we cross-identified
individual model components along the jet using their distance from the VLBI
core, flux density and size. To test the stability of the VLBI core position
with time and between different frequencies we compared the core separations of
the jet components of each epoch with respect to the adjacent observations, and
we found no systematical position offsets. Supported by a graphical
analysis, which is presented in Figure~1, we obtained a satisfactory
identification scheme for the kinematics in the jet of 0716+714. This scenario
consists of 11 superluminal components moving linearly away from the core.
\subsection{The New Kinematical Model}
The components in this scenario move with 0.3\,mas\,yr$^{-1}$ to
0.5\,mas\,yr$^{-1}$ in the inner part of the jet ($r\leq3$\,mas) and
with up to 0.8\,mas\,yr$^{-1}$ in the outer regions. Assuming $z=0.3$ for
0716+714, the measured angular separation rates correspond to speeds of
4.0\,$c$ to 14.3\,$c$ (where we adopt $H_0=70$\,km\,s$^{-1}$\,Mpc$^{-1}$ and
$q_0=0.5$). A previous preliminary analysis by Tian et al. (2001), 
who used only part of the data presented here, gave very similar speeds. The
speeds found in this paper are lower than the 15\,$c$ to 20\,$c$ found by
Jorstad et al. (2001), who observed the source only over a short time interval
(3 yr). The difference between their and our result may be explained 
by slightly different (and unfortunately not unambiguous) map parameterizations
using different Gaussian models. With an average apparent jet speed of about
7\,$c$ and possibly higher speeds to up to 20\,$c$, 0716+714 is considerably
faster than a `typical' BL\,Lac object, for which speeds of $\leq5\,c$ are
regarded as normal (Gabuzda et al. 2000 and references therein).
\subsection{Kinematics and Geometry}
Using the measured motion of C3 (10.3\,$c$ for $z=0.3$), the fastest, well
constrained component in our model, we can place limits on the kinematics and
geometry of 0716+174. Adopting $\beta_{\rm app}=\frac{\beta \sin\theta}{1-\beta
\cos\theta}$ we find for the jet viewing angle $\theta$, which maximizes the
apparent speed a value of $\theta_{\rm max} =5.6^\circ$. The minimum Lorentz factor
$\gamma_{\rm min} = ( 1 + \beta_{\rm app}^2)^{1/2}$ then is $10.3$, which
corresponds to a Doppler factor of $\delta=\gamma^{-1}(1 - \beta
\cos\theta_{\rm max})^{-1} = [\gamma-\sqrt{\gamma^2-1}\,\cos\theta_{\rm max})]^{-1} =
10.3$. For smaller viewing angles the Doppler-factor increases and reaches its
maximum of $\delta_{max} = 2 \gamma$ at $\theta \rightarrow 0$. For C3 this
yields $\delta_{max} =20.6$. To explain the large range of observed apparent
speeds (see Figure~1) as an effect of spatial jet bending, a Lorentz factor of
$\gamma>15$ and a viewing angle of the VLBI jet of $\theta <2^\circ$ are more
likely. Under these circumstances, the Doppler factor would be $\delta>30$.

Such high Doppler factors are in good agreement with those derived from intrinsic
intraday variability at cm-wavelengths, which are required to bring the high
apparent brightness temperatures of up to $10^{17}$\,K down to the
inverse-Compton limit of $10^{12}$\,K.

\acknowledgements
We thank S. Jorstad and A. Marscher, the group of the VLBA 2cm Survey and the
group of the CJF-Survey for providing their data. This work made use of the
VLBA, which is an instrument of the National Radio Astronomy Observatory, a
facility of the National Science Foundation, operated under cooperative
agreement by Associated Universities, Inc. and the European VLBI Network, which
is a joint facility of European, Chinese, South African and other radio 
astronomy institutes funded by their national research councils.



\begin{references}
\reference Britzen, S., et al.\ 1999, In ASP Conf. Ser. Vol. 159: BL Lac
Phenomenon, ed. L.O. Takalo \& A. Sillanp\"a\"a, (San Francisco: ASP), 431

\reference Eckart, A., et al.\ 1986, \aap, 168, 17

\reference Eckart, A., et al.\ 1987, \aaps, 67, 121

\reference Gabuzda, D.C., et al.\ 1998, \aap, 333, 445

\reference Gabuzda, D.C., Pushkarev, A.B., \& Cawthorne, T.~V.\ 2000, \mnras,
319, 1109 

\reference Heeschen, et al.\ 1987, \aj, 94, 1493  

\reference Jorstad, et al.\ 2001, \apjs, 134, 181 

\reference Kellermann, K.I., et al.\ 1998, \aj, 115, 1295 

\reference Quirrenbach, A., et al.\ 1991, \apjl, 372,
L71 

\reference Ros, E., et al.\ 2001, \aa, 376, 1090

\reference Schalinski, C.J., et al.\ 1992, in Variability of Blazars, ed. E.
Voltaoja \& M. Valtonen (Cambridge University press), 225 

\reference Tian, W.W., et al.\ 2001, IAU Symposium 205, Galaxies and their
Constituents at the Highest Angular Resolutions, Manchester, ed. R.T. Schilizzi,
S.N. Vogel, F. Paresce \& M.S. Elvis, 96

\reference Wagner, S.J., et al.\ 1996, \aj, 111, 2187

\reference Witzel, A., et al.\ 1986, Mitteilungen der Astronomische Gesellschaft, 65, 239 

\reference Witzel, A., et al. \ 1988, \aap, 206, 245 

\reference Zensus, J.A., et al.\ 2002, \aj, 124, 662

\end{references}
\end{document}